\documentclass[aps,prapplied,superscriptaddress,reprint]{revtex4-2}

\usepackage{amsmath,amsfonts,amssymb}
\usepackage{graphicx}
\usepackage{hyperref}

\begin{document}
\title{Single-laser feedback cooling of optomechanical resonators}  

\author{Arvind Shankar Kumar}
\affiliation{Department of Physics and Nanoscience Center, University of Jyv{\"a}skyl{\"a}, P.O. Box 35, FI-40014 University of Jyv{\"a}skyl{\"a}, Finland}

\author{Joonas Nätkinniemi}
\affiliation{Department of Physics and Nanoscience Center, University of Jyv{\"a}skyl{\"a}, P.O. Box 35, FI-40014 University of Jyv{\"a}skyl{\"a}, Finland}

\author{Henri Lyyra}
\affiliation{Department of Physics and Nanoscience Center, University of Jyv{\"a}skyl{\"a}, P.O. Box 35, FI-40014 University of Jyv{\"a}skyl{\"a}, Finland}

\author{Juha T. Muhonen}
\email{juha.t.muhonen@jyu.fi}
\affiliation{Department of Physics and Nanoscience Center, University of Jyv{\"a}skyl{\"a}, P.O. Box 35, FI-40014 University of Jyv{\"a}skyl{\"a}, Finland}

\begin{abstract}
    Measurement-based control has emerged as an important technique to prepare mechanical resonators in pure quantum states for applications in quantum information processing and quantum sensing. Conventionally this has required two separate channels, one for probing the motion and another one acting back on the resonator.
    In this work, we analyze and experimentally demonstrate a technique of single-laser feedback cooling, where one laser is used for both probing and controlling the mechanical motion. We show using an analytical model and experiments that feedback cooling is feasible in this mode as long as certain stability requirements are fulfilled.
    Our results demonstrate that, in addition to being more experimentally feasible construction, the interference effects of the single-laser feedback can actually be used to enhance cooling at some parameter regimes.
\end{abstract}

\maketitle

\section{Introduction}
Micro- and nanoscale mechanical resonators have emerged as an important tool for various applications in quantum information processing \cite{Aspelmeyer2014,laucht_roadmap_2021} and quantum sensing \cite{degen_quantum_2017}. One main motivation lies in the promise of using the mechanical resonators as quantum transducers between different quantum systems as the mechanical degree of freedom is easy to couple to other systems, including optical fields and different types of qubits \cite{Geller2005,Stannigel2011,Ovartchaiyapong2014,Norte2018}. Another interesting aspect are the fundamental questions on quantum decoherence of massive objects, and other quantum phenomena \cite{wollman_quantum_2015,pirkkalainen_squeezing_2015,lecocq_quantum_2015,chu_creation_2018,ockeloen-korppi_stabilized_2018,riedinger_remote_2018} at macroscopic scale.

One major challenge in the use of mechanical resonators in quantum applications is their coupling to the thermal environment, which acts as a source of classical noise, represented by the finite average phonon number of the resonator and its variance. This classical noise can obscure the quantum information that is imprinted on the resonator. Thus, it is an important challenge to cool these resonators close to their motional quantum ground state, and minimize the phonon noise in the system. This was achieved using sideband-cooling about a decade ago \cite{oconnell_quantum_2010,chan_laser_2011,teufel_sideband_2011}.

Somewhat counter-intuitively the cooling can also be achieved with measurement based feedback, similarly as in classical systems, even down to the quantum ground state \cite{cohadon_cooling_1999,wilson_measurement-based_2015}. Indeed, in the past few years, measurement-based feedback cooling has emerged as an important technique for achieving cooling of mechanical resonators into the quantum regime \cite{Zhang2017,Rossi2018,Guo2019,Borrielli2021} (these methods have also been applied to qubits \cite{vijay_stabilizing_2012}). In this technique, the measured displacement of the oscillator is fed back into the sample as a force modulation that then allows for damping of the oscillator \cite{Rossi2018}. Demonstrations have managed to cool mechanical resonators to the quantum ground state at cryogenic temperatures \cite{Rossi2018} and come close to it even at room temperature \cite{Guo2019}. 

As opposed to sideband cooling \cite{Aspelmeyer2014,Teufel2011}, measurement-based feedback cooling is performed on optomechanical systems in the non-resolved sideband regime $\kappa\gg\omega_M$ (where $\kappa$ is the optical decay rate of the optical cavity and $\omega_M$ is the natural frequency of the mechanical mode of interest) as in this regime the readout optical field responds instantaneously to changes in the mechanical motion. Usually the displacement of the resonator is measured using a balanced homodyne interferometer and then the output of the homodyne detector - after applying various filters - is fed into an auxiliary laser or a piezo-stage in order to provide the force modulation. 
 
Here we study the case where only one laser beam is used for both probing and inducing the feedback force \cite{Habibi2016,Guo2019} to the optomechanical system, in spirit similar to existing literature about ''in-loop'' light \cite{Zippilli2018}. The single-laser technique holds much promise due to its simplicity and efficiency of implementation. However, one could naively expect that the feedback modulation of the laser intensity at the mechanical resonance frequency would show up as a component of the homodyne signal without being transduced through the displacement of the resonator. This would then in turn affect the displacement measurement, which could then affect the feedback force and a vicious cycle could be born. This raises issues of both the ability to achieve significant feedback cooling and being able to interpret the homodyne signal as a readout of the mechanical displacement of the resonator. Given these considerations, it is important to investigate the technique and the interpretations that can be drawn from it in detail.
In this work, we analyze the single-laser feedback cooling through a classical analytical model - which is then validated through a numerical study - and further implement such a setup experimentally confirming the model.

Through our model we find that while the interference between measurement and feedback modulation in the single-laser setup can cause the system to go to an unstable regime for large enough feedback gain $\gamma_{fb}$ and/or deviation from the typically used homodyne angle ($\phi$) of $\pi/2$ (where the homodyne signal is directly proportional to the resonator displacement), there is also a region of parameter space $(\gamma_{fb},\phi)$ where the system is stable. Within this stable region, the spectrum of the steady-state homodyne signal around $\omega_M$ is still always proportional to the displacement of the oscillator for any feedback phase $\theta_{fb}$ and homodyne angle $\phi$. Thus the displacement readout mechanism is preserved, although interference effects between the modulated signal and the readout signal are found to cause a change in the dynamics of the resonator from what is expected purely without such interference (as in the auxiliary laser case), and also result in a feedback gain-dependant transduction between the homodyne signal and resonator displacement. We further find that, by operating close to the instability point, it is possible to use these interference effects to achieve a damping rate of the resonator that is higher than what can be achieved using an auxiliary laser, and the improvement in damping rate compared to the auxiliary laser case improves with increasing feedback gain. In addition to this, we are also able to compare our analytical and numerical model to experimental results over the whole range of feedback phases, and use the model to extract resonator displacement and cooling with applied feedback gain. However, further studies will be required to estimate the impact of noise and the ultimate cooling limits in the single-laser case. Our model is also fully classical and does not include quantum effects. Nevertheless, we believe its simplicity will aid in the design of optomechanical feedback loops. In addition, our model demonstrates that when using the same laser for both feedback and measurement, care must be taken in interpreting the results as the transduction factor from mechanical motion to homodyne signal will be affected.

\section{Setup}
\begin{figure}[tb]
    \centering
    \includegraphics[width=0.5\textwidth]{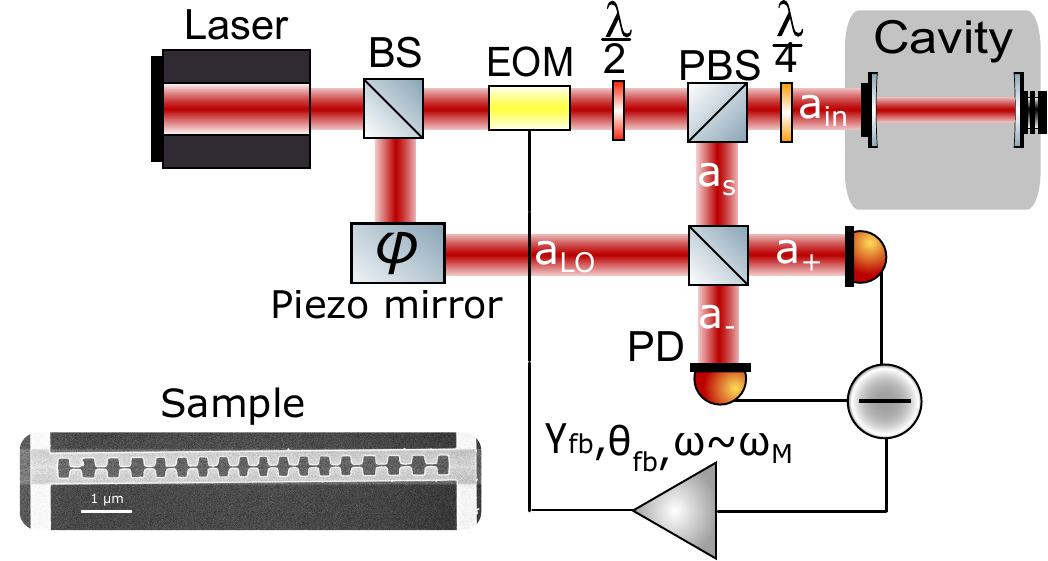}
    \caption{Schematic of the measurement setup and scanning electron microscope image of the silicon nanobeam sample.}
    \label{fig:setup}
\end{figure}

The analyzed setup is presented in Fig. \ref{fig:setup}. It is a balanced homodyne interferometer setup, with an added control loop for the feedback cooling. In the measurement arm, laser light incident on the sample interacts with the mechanics through radiation pressure coupling and the reflected light from the sample then encodes the information on the mechanical position. This reflected light (called the signal (S) branch) is then interfered with the other arm of the homodyne interferometer (the local oscillator (LO) branch) - which is also from the same laser source, but has a phase difference of $\phi$ imprinted on it - at a 50:50 beamsplitter and the two outputs of said interference (labelled + and -) are collected at two photo-detectors. A single final output is obtained by subtracting the photocurrents at the two detectors.

To implement single-laser feedback cooling, the output from the photodetector, in addition to being read out by a spectrum analyser, is digitized and multiplied in the frequency domain with a filter function that includes a band-pass filter, a gain factor, and a phase factor. The modified signal is then converted back into a time domain voltage signal that is fed into an electro-optical-modulator (EOM) that modulates the laser power incident on the mechanical resonator.

\section{Analytical Model}
In order to build a model for the single-laser feedback, we start with the signal from the balanced homodyne detector $H$. We assume the two beams are linearly polarized coherent states with complex parameters $a_\mathrm{s}$ and $a_\mathrm{lo}$ and the homodyne signal can then be written as
\begin{equation}
    H=\lvert a_+ \rvert^2 - \lvert a_- \rvert^2 = a_s^*a_{LO}-a_{LO}^*a_s.
\end{equation}
We then assume input-output relationship for our system $a_s=\sqrt{\kappa_{ex}}a$, where $\kappa_{ex}$ is the optical decay rate towards the detector of the cavity mode $a$ \cite{note1}, and a cavity mode that is detuned by the mechanical motion. We can then derive $a(t)$ and the expected homodyne signal for a given input amplitude $a_{in}$ (where $\lvert a_{in} \rvert^2 = P_{in}/(\hbar \omega_L) $ represents the rate of incoming photons and $P_{in}$ is the input laser power at the laser frequency $\omega_L$) in the non-resolved sideband regime ($\kappa \gg \omega_M$) (where the optical mode reaches a steady state much faster than the mechanics) \cite{Aspelmeyer2014}. Hence, we use
\begin{equation}
    a(t)=\frac{\kappa_{ext}a_{in}}{-i\Delta(t)+\kappa/2},
\end{equation}
giving
\begin{equation}
     H=\lvert a_{LO}\rvert \lvert a_{in}\rvert 4 \eta \frac{\cos{\phi}+\Delta_n\sin{\phi}}{1+\Delta_n^2},
     \label{Heqnl}
\end{equation}
where $\eta = \kappa_{ex}/\kappa$, $\Delta_n = 2\Delta(t)/\kappa$ and the effective detuning $\Delta(t)=\Delta_0 + g_0 x(t)$, where $g_0$ is the optomechanical coupling rate, $\Delta_0$ is the cavity detuning, and $x(t)$ is the displacement of the mechanical resonator, normalized with respect to $x_{zpf}$, the zero-point motion of the resonator.
When the resonator displacement is small enough so that $\Delta_n\ll1$, 
\begin{equation}
     H\approx\lvert a_{LO}\rvert \lvert a_{in}\rvert 4 \eta(\cos{\phi}+\Delta_n\sin{\phi}).
     \label{Heq}
\end{equation}
For our analysis, we work in this regime where the homodyne signal is linearly related to the resonator displacement (in the absence of feedback effects).

The signal from the homodyne detector is then fed into a bandpass filter, which is set up to pass through only frequencies close to the natural frequency of the resonator $\omega_M$, in addition to scaling the signal by a controllable feedback gain $\gamma_{fb}$ and adding a feedback phase $\theta_{fb}$. This is in-turn fed to the EOM which modulates the initial input signal intensity and feeds it back into the sample. The feedback phase $\theta_{fb}$ (which represents the phase difference with which the feedback force is applied on the sample) can be tuned so that a damping or driving force is applied that results in cooling or heating of the resonator respectively. In the conventional auxiliary laser case, $\theta_{fb}=\pi/2$ is the optimal cooling point and $\theta_{fb}=-\pi/2$ is the optimal heating point, but as we will see in this work, this does not always hold true for the single-laser case.

If we consider each loop of feedback as a separate process where the input to the sample changes, we can describe it as 
\begin{equation}
    \lvert a_{in}^{n+1} \rvert^2 =\lvert a_{in}^n \rvert^2(1 + \gamma_c H_{EOM}^{n'}),
    \label{aineq}
\end{equation}
where $\lvert a_{in}^n \rvert^2$ is the modulated input intensity at the $n$\textsuperscript{th} loop, $\gamma_c=4 \eta(V_{detector}/V_{EOM})(\lvert a_{LO}\rvert/\lvert a_{in}^0 \rvert)$ where $V_{detector}$ and ${V_{EOM}}$ are the detector and EOM optical power to voltage conversion parameters, and  $H_{EOM}^{n'}=H_{EOM}^{n}/4\eta\lvert a_{LO}\rvert \lvert a_{in}^0\rvert$ where
\begin{equation}
    H_{EOM}^{n}=\gamma_{fb}e^{i\theta_{fb}} H_n^{\omega\sim\omega_M}
    \label{EOMeq}
\end{equation}
is the electrical input signal to the EOM at the $n$\textsuperscript{th} loop with feedback gain $\gamma_{fb}$ and phase $\theta_{fb}$. We mark as $H_n^{\omega\sim\omega_M}$ the homodyne signal (at the $n$\textsuperscript{th} loop) that has been filtered to include only frequencies around $\omega_M$. 

Starting from an unmodulated input of  $\lvert a_{in}^0 \rvert^2$, we can then iteratively solve Eqs.~(\ref{Heq}), (\ref{aineq}) and (\ref{EOMeq}) and derive a general equation for the homodyne signal after an arbitrarily large N number of feedback loops $H_s^{\omega\sim\omega_M}$ (See section A of Supplemental Material (SM) for derivation) 
\begin{equation}
\begin{split}
    H_s^{\omega\sim \omega_M}=\lvert a_{LO}\rvert \lvert a_{in}^0 \rvert 4 \eta\sin{\phi}\frac{2g_0}{\kappa}\\\times\lim_{N\to\infty} [\sum_{n=0}^{N}x_{N-n}(t)(\frac{1}{2}\gamma_c e^{i\theta_{fb}}\gamma_{fb}\cos{\phi})^n],
    \label{HNeq}
\end{split}
\end{equation}
where $x_n(t)$ is the amplitude of the mechanical oscillation for the $n$\textsuperscript{th} feedback loop. Note that we do not explicitly model the effects of the light field to the mechanical displacement at this point, we merely assume that at each feedback loop the mechanical resonator displacement is $x_n(t)$, that can differ at each loop. From this equation, we see that the history of the resonator displacements at each feedback loop step is carried over into the signal through the modulations of the laser intensity that they result in. For large number of time steps, the series terms with displacements corresponding to older time steps start to die out when the gain and phase factors are small enough, and the series in Eq. (\ref{HNeq}) converges. The convergence of the series at all feedback phases requires that the feedback gain is small enough so that the term raised to the power of $n$ decays faster than the amplitude grows when the feedback heats up the resonator.

Assuming that the resonator displacement converges to $x_s(t)=\lim_{n\to\infty} x_n(t)$, the feedback-modulated input signal and corresponding homodyne signal around $\omega_M$ can be written as 
\begin{equation}
\begin{split}
\lvert a_{in}^s \rvert^2 = \lvert a_{in}^0 \rvert^2 [1+\gamma_c e^{i\theta_{fb}}\sin{\phi}\gamma_{fb} \frac{2g_0}{\kappa}x_s(t)\\ \times\sum_{n=0}^{\infty}(\frac{1}{2}\gamma_c e^{i\theta_{fb}}\gamma_{fb}\cos{\phi})^n] 
\end{split}
\label{anseq}
\end{equation}
and
\begin{equation}
\begin{split}
    H_s^{\omega\sim\omega_M} = \lvert a_{in}^0 \rvert\lvert a_{LO}\rvert 4 \eta\sin{\phi}\frac{2g_0}{\kappa}x_s(t)\\
    \times\sum_{n=0}^{\infty}(\frac{1}{2}\gamma_c e^{i\theta_{fb}}\gamma_{fb}\cos{\phi})^n,
\end{split}
\label{Hseq}
\end{equation}
We note that the series term in Eqs. (\ref{anseq}) and (\ref{Hseq}) correspond to interference effects due to the feedback modulations (used to control the resonator) showing up in the homodyne signal. As seen in these equations, these interference effects die out when $\phi$ is tuned exactly to $\pi/2$.

In the case where the resonator displacement converges to $x_s(t)$, the homodyne signal $H_s^{\omega\sim\omega_M}(t)$ remains proportional to the $x_s(t)$ despite the interference effects of the feedback, preserving the displacement readout mechanism. However, the interference effects result in a modification of the transduction parameter $\beta=H_s^{\omega\sim\omega_M}/x_s$, which is now given by
\begin{equation}
    \beta=\beta_{\pi/2}\sin{\phi}\sum_{n=0}^{\infty}(\frac{1}{2}\gamma_c e^{i\theta_{fb}}\gamma_{fb}\cos{\phi})^n
    \label{transduction}
\end{equation}
where $\beta_{\pi/2}=\lvert a_{in}^0 \rvert\lvert a_{LO}\rvert 4 \eta\frac{2g_0}{\kappa}$ is the transduction parameter at $\phi=\pi/2$. 
Similarly, the steady state input signal $a_{in}^s$ also contains a feedback component that is proportional to $x_s$, with the interference effects again modifying the proportionality relation.

To model the effect of this feedback-modulated input (Eq. (\ref{anseq})) on the resonator, we consider the linear response regime for the mechanical resonator treated as a damped harmonic oscillator. The resonator displacement in the frequency domain is then given by $x_{s}(\omega)=\chi_{eff}(\omega)(F^0_{rp}+ F_{th})$, where $F^0_{rp}=4\eta\frac{g_0}{\kappa}\lvert a_{in}^0\rvert^2$ is the radiation pressure force from the unmodulated input and $F_{th}$ is the thermal driving force, and the feedback-modulated mechanical susceptibility $\chi_{eff}(\omega)$ is given by
\begin{equation}
    \chi_{eff}(\omega)=\frac{\chi_0(\omega)}{1-\chi_0(\omega)h_{fb}^{eff}(\gamma_{fb},\theta_{fb})},
    \label{xseq}
\end{equation}
where $\chi_0(\omega)=\omega_M/(\omega_M^2-\omega^2-i\Gamma\omega)$ is the bare mechanical susceptibility of the resonator.
Here, $h_{fb}^{eff}(\gamma_{fb},\theta_{fb})$ is the effective filter function of this idealized bandpass filter (BPF) that takes into account the interference effects from the single-laser feedback
\begin{equation}
\begin{split}
h_{fb}^{eff}(\gamma_{fb},\theta_{fb})=K \frac{2g_0}{\kappa}e^{i\theta_{fb}}\sin{\phi}\gamma_c\gamma_{fb} \\
\times [\sum_{n=0}^{\infty}(\frac{1}{2}\gamma_c e^{i\theta_{fb}}\gamma_{fb}\cos{\phi})^n],   
\end{split}
\end{equation}
where $K= \lvert a_{in}^0\rvert^2 (4 \eta g_0/\kappa)$.

From this equation of resonator displacement, we can now extract the effective damping rate, which determines the heating/cooling rate, in the presence of feedback. This is given by
\begin{eqnarray}
\Gamma^{eff}&=&\Gamma+K \gamma_c \sin{\phi}\gamma_{fb} \frac{2g_0 }{\kappa}
\nonumber \\
&\times &\Im(e^{i\theta_{fb}}[\sum_{n=0}^{\infty}(\frac{1}{2}\gamma_c e^{i\theta_{fb}}\gamma_{fb}\cos{\phi})^n]),
\label{gamma}
\end{eqnarray}
where $\Im(..)$ represents the imaginary part operation.

\begin{figure*}
    \centering
    \includegraphics[width=1\textwidth]{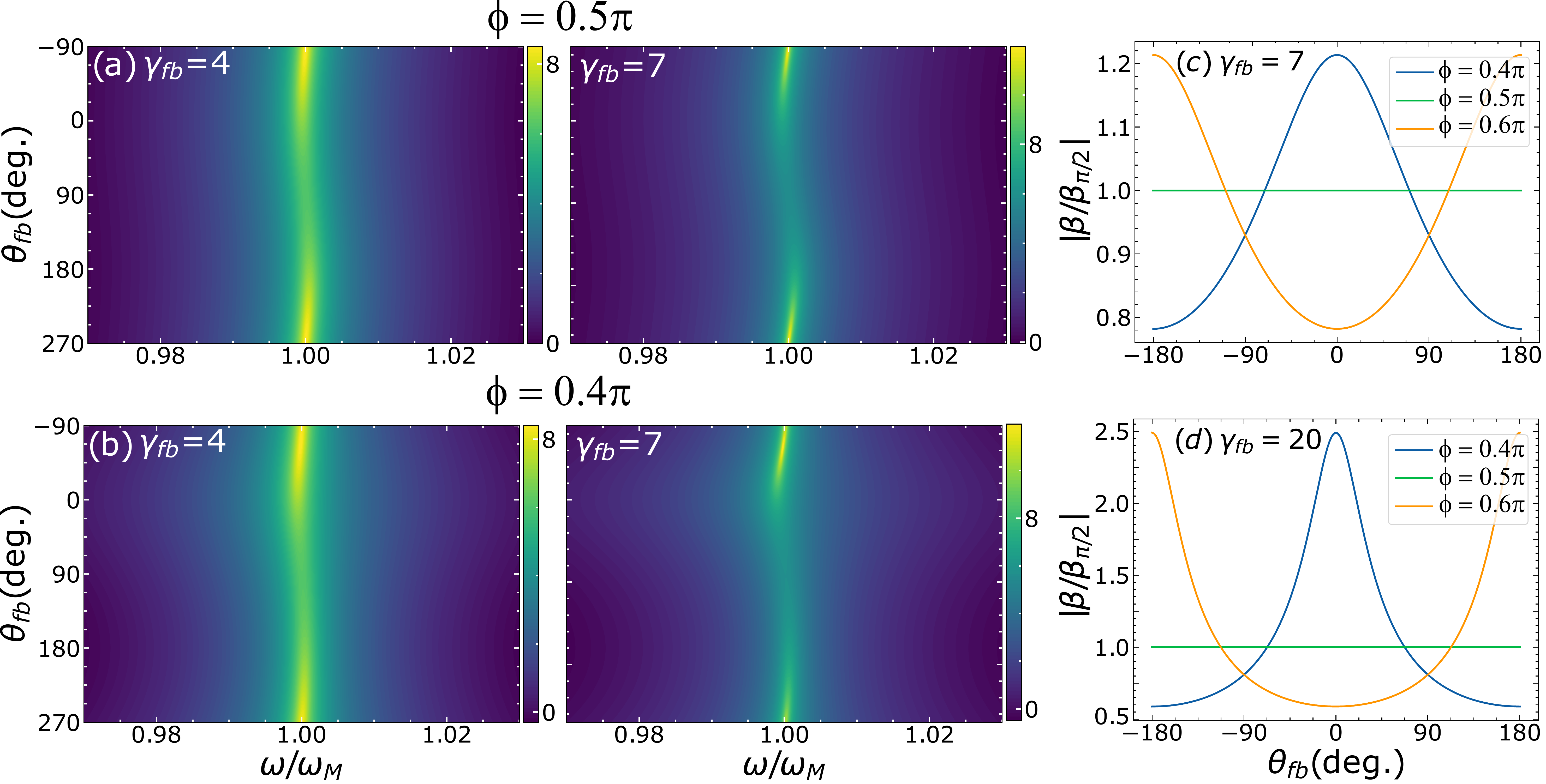}
    \centering
    \caption{(a),(b) Calculated feedback phase dependence of homodyne signal spectrum from analytical model for $\kappa=200$ GHz, $g_0=8$ MHz, $\eta=0.01$, $\Gamma=15$ kHz, $\omega_M=8$ MHz, $P_{in}=10$ \textmu W, $\frac{\lvert a_{LO}\rvert}{\lvert a_{in}^0 \rvert}=10$, $\gamma_c \simeq 0.6\frac{\lvert a_{LO}\rvert}{\lvert a_{in}^0 \rvert}4 \eta$  and feedback gains (left) $\gamma_{fb}=4$ and (right) $\gamma_{fb}=7$ for homodyne angles $\phi=\pi/2$ and $\phi=0.4\pi$ respectively. The color bar represents the PSD of the calculated homodyne signal in an arbitrary logarithmic scale. (c),(d) Transduction parameter $\beta$ (as a ratio of $\beta_{\pi/2}$) plotted versus $\theta_{fb}$ for $\phi=0.4 \pi,\pi/2, 0.6\pi$ for feedback gains $\gamma_{fb}=7$ and $\gamma_{fb}=20$ respectively.}
    \label{model scans}
\end{figure*}

\begin{figure}
    \centering
    \includegraphics[width=0.5\textwidth]{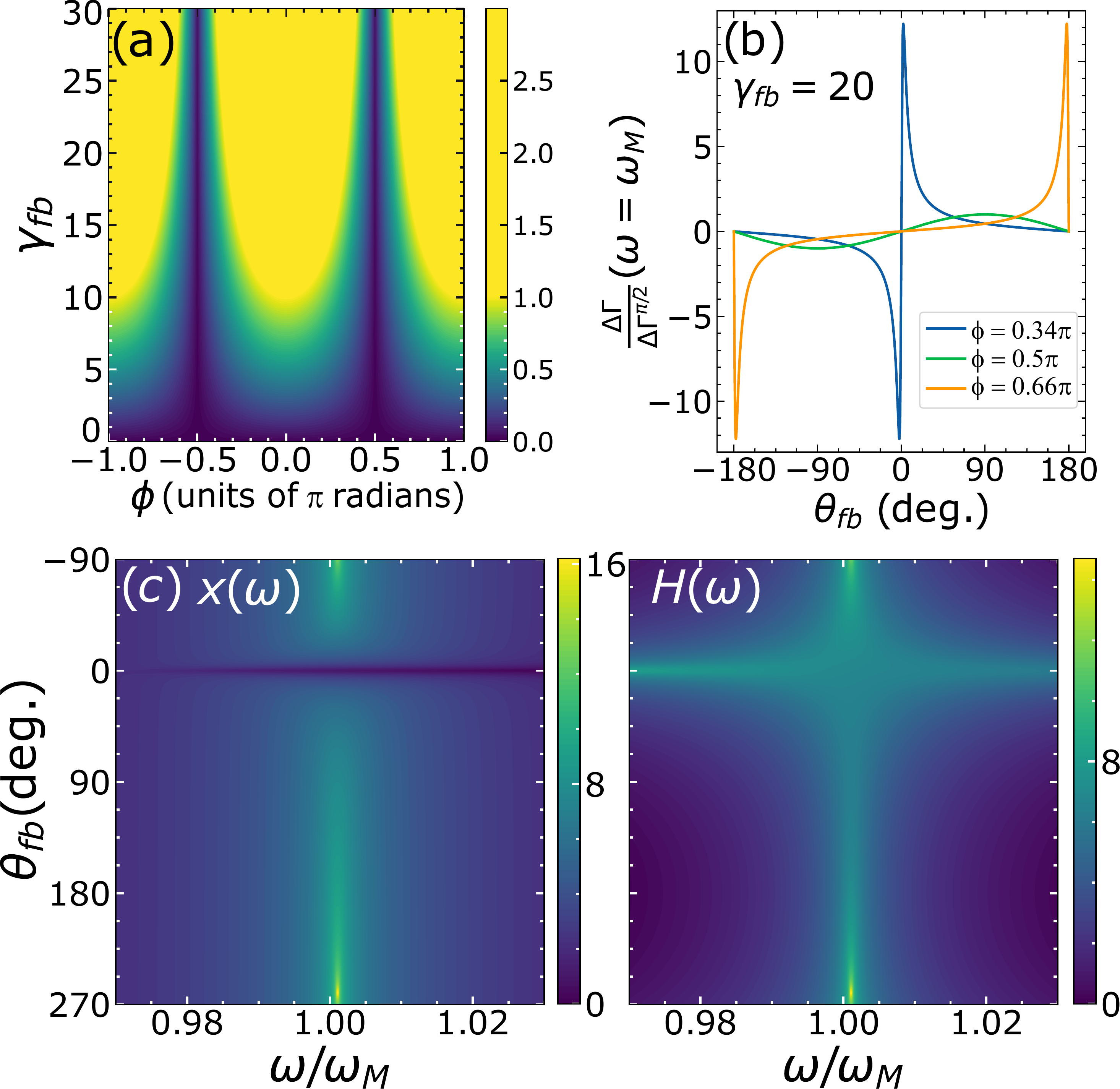}
    \caption{(a) The stability condition expression (given in Eq. (\ref{stabilityeq})) plotted versus $\gamma_{fb}$ and $\phi$. The yellow regions represent regions of instability. (b) The change in effective damping rate $\Delta\Gamma=\Gamma^{eff}-\Gamma$ (given in Eq. (\ref{gamma})) at $\omega=\omega_M$ plotted versus $\theta_{fb}$ for $\phi=0.34\pi,\pi/2,0.66\pi$, for $\gamma_{fb}=20$, normalized with respect to $\Delta\Gamma_0=\Delta\Gamma(\theta_{fb}=\pi/2,\phi=\pi/2)$ (c) Calculated feedback phase dependence of resonator displacement spectrum (left) and homodyne signal spectrum (right)  from analytical model for $\phi=0.34 \pi$ and $\gamma_{fb}=20$. The color bar represents the corresponding PSD in an arbitrary logarithmic scale. Notably the mechanical resonator is cooled at $\theta_{fb} = 0$, but this does not appear in the measurement signal.}
    \label{stability}
\end{figure}

In our experiments, we measure and plot the frequency spectrum of the homodyne signal (for different applied feedback gain), which is used to infer the resonator displacement, as a function of the feedback phase. This is done to understand the behavior of the resonator over the full range of feedback phases, and to locate the optimal cooling point. To furnish comparisons to this experiment, in Fig. \ref{model scans} we plot the predicted homodyne signal spectrum from our model  as a function of feedback phase for two $\gamma_{fb}$ values corresponding to feedback gains in range of our experiment, for $\phi=\pi/2$ and  $\phi=0.4\pi$ (i.e 20 \% deviation from $\pi/2$ ), and system parameters (given in figure caption) similar to those achievable in our experiment. We estimate $\gamma_c \simeq 0.6\frac{\lvert a_{LO}\rvert}{\lvert a_{in}^0 \rvert}4 \eta$ (using the estimates of conversion factors from EOM and photo-detector properties). It is to be noted that, while we use $\eta=0.01$ for the plots in Fig. \ref{model scans}, the results are similar for all $\eta$ apart from changes in the values of the feedback gain required to see similar results. This is due to the fact that we do not consider noise and other limiting factors for cooling here.

For both $\phi=\pi/2$ (where the interference effects in Eq. (\ref{Hseq}) die out) and  $\phi=0.4\pi$, we see similar regions of strong and weak homodyne signal at $\theta_{fb}=-\pi/2$ and $\pi/2$, respectively corresponding to heating and cooling of the mechanical resonator, similar to what is expected for the auxiliary laser case. The comparison of these predicted homodyne signals to the corresponding resonator displacement is shown in section B of SM. It can be seen that, when $\phi$ is in this range, the interference effects are small enough so that the homodyne signal plots in Fig. \ref{model scans} can be directly used to infer the approximate behavior of the resonator displacement as a function of the feedback phases. However, the feedback phase dependence of the transduction parameter between the homodyne signal and resonator displacement (Eq. (\ref{transduction})) will still be needed to considered to infer the exact displacement when $\phi\neq\pi/2$. 

Outside this range of ($\phi,\gamma_{fb}$), even the qualitative behavior of homodyne signal and resonator displacement start to look different in their feedback phase dependence and care has to be taken when trying to infer the resonator displacement from the homodyne signal using such plots in these cases (see Fig. \ref{stability}(c)). The feedback phase dependence of the homodyne signal and corresponding resonator displacement also start to differ significantly from what is seen in the $\phi=\pi/2$ case. However, this regime - where the interference effects play a dominant role - is also interesting because, it is seen that the interference effects modify the resonator dynamics sufficiently so that the maximum feedback cooling achieved is larger than for the 'optimal' case of $\phi=\pi/2$.

In order to further explore these effects, we return to our analytical model where, from Eqs. (\ref{anseq}) and (\ref{Hseq}), we see that the interference term diverges unless a stability condition
\begin{equation}
    \frac{1}{2}\gamma_c\gamma_{fb}\lvert\cos{\phi}\rvert < 1,
    \label{stabilityeq}
\end{equation}
is satisfied. Hence, the system is stable for small enough feedback gain or deviation from $\phi=\pi/2$. For larger gain or deviation in $\phi$, the interference effects drive the system into an unstable regime which results in uncontrolled resonator heating at all feedback phases. The stability condition expression given in Eq. (\ref{stabilityeq}) is plotted in Fig. \ref{stability}(a) versus ($\phi,\gamma_{fb}$). The yellow regions represent regions of instability. It can be seen that these unstable regions start out at finite $\gamma_{fb}$ at $\phi=0,\pi$ and expand over larger range of $\phi$ as $\gamma_{fb}$ is increased, while $\phi=\pm \pi/2$ always remains a stable point. 

In order to investigate the effect of these interference terms on resonator dynamics, specifically regarding cooling, we plot the change in effective damping rate $\Delta\Gamma=\Gamma^{eff}-\Gamma$ from Eq. (\ref{gamma}) at $\omega=\omega_M$ versus $\theta_{fb}$ at different $\phi$ values, for a given $\gamma_{fb}$ (as seen in Fig. \ref{stability}(b) ). We see that the effective damping rate, and hence the cooling rate, can be enhanced to be larger than what is achievable in the $\phi=\pi/2$ case where the interference effects die out. We find that, for a given value of $\gamma_{fb}$, the maximum cooling rate achievable (by tuning $\theta_{fb}$) increases as $\phi$ approaches the instability regions. However, it is also observed that the cooling regions become narrower in $\theta_{fb}$ (seen in Fig. \ref{stability}(b), for example) as we approach the instability, thus making them more difficult to resolve. Additionally, it is to be noted that a determination of the actual cooling limits requires a treatment of the noise in the system, and this is left as subject of future work. However, we are here able to propose a mechanism to enhance cooling rates beyond what is conventionally possible in the auxiliary laser case. 

Finally, care has to be taken while interpreting the homodyne signal in this region of parameter space, since the changes in transduction parameter with $\theta_{fb}$ become significant as we approach the instability regions. For example, we see in Fig. \ref{stability}(c) that the region of cooling seen in the resonator displacement plot (left) (close to $\theta_{fb}=0$) is not seen in the homodyne signal (right). Thus, it is essential to interpret the actual resonator displacement from the homodyne signal through the transduction parameter derived in the model in this case, or use alternative methods of reading out the effective temperature like an out-of-loop measurement \cite{Rossi2018}.

\section{Numerical Study}
To obtain further insight into the validity of our analytical model, we can numerically solve the response of the resonator to a modulated input assuming a realistic filter function of the BPF and directly considering the full response of the resonator to the modulated input. 

To this end, we consider the modulated input in frequency space.
\begin{equation}
\lvert a_{in}(\omega) \rvert^2 =\lvert a_{in}^0\rvert^2(\delta(\omega) + \gamma_c H_{EOM}^{'}(\omega))
\label{ainnumeq}
\end{equation}

In this case, $H_{EOM}(\omega)$ is derived from the actual filter function of the BPF as
\begin{equation}
    H_{EOM}(\omega)=\gamma_{fb}e^{i\theta_{fb}}h_{BPF}(\omega)H(\omega)
    \label{HEOMnumeq}
\end{equation}
where $h_{BPF}$ is a Lorentzian of given bandwidth centered at $\omega_M$, and $H_{EOM}^{'}(\omega)$ is normalized in a similar manner as in the previous section. $H(\omega)$ depends on the dynamics of the resonator in frequency space $x(\omega)$ through taking the Fourier transform of Eq. (\ref{Heq}), given a modulated input signal, resulting in 
\begin{equation}
    H=\lvert a_{LO}\rvert 4 \eta (\lvert a_{in}(\omega)\rvert \cos{\phi}+\lvert a_{in}(\omega)\rvert*\Delta_n(\omega)\sin{\phi})
    \label{Hnumeq}
\end{equation}
where * is the convolution operation. 
The resonator dynamical response in frequency space to the modulated input is then given by
\begin{equation}
x(\omega)=\chi_0(\omega)*(F_{rp}(\omega)+F_{th}(\omega))
\label{xnumeq}
\end{equation}
where $F_{rp}(\omega)=F_{rp}^0 (1+\gamma_c H_{EOM}(\omega))$ is the radiation pressure force due to the modulated input. 
  
\begin{figure}
    \centering
    \includegraphics[width=0.5\textwidth]{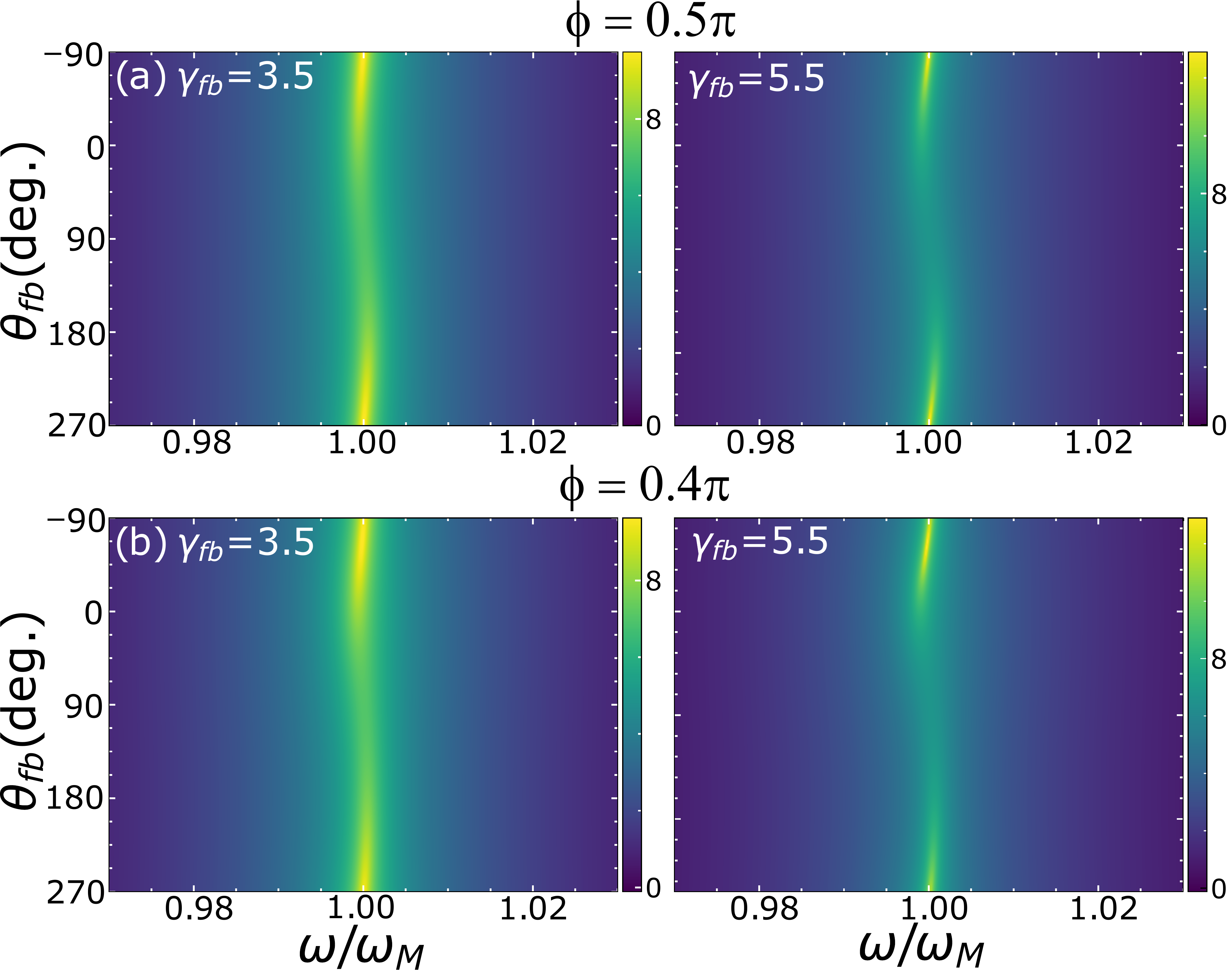}
    \caption{Numerical model results for the feedback phase dependence of homodyne signal spectrum, simulated for a BPF bandwidth of 80 kHz and the same system parameters given in the caption to Fig. \ref{model scans} and numerical model feedback gains (left) $\gamma_{fb}=3.5$ and (right) $\gamma_{fb}=5.5$ (feedback gains for the numerical model are further scaled by a factor of $100$), for (a) $\phi=\pi/2$ and (b) $\phi=0.4\pi$. The color bar represents the PSD of the simulated homodyne signal in an arbitrary logarithmic scale.}
    \label{numerical scans}
\end{figure}

Eqs. (\ref{ainnumeq}), (\ref{HEOMnumeq}), (\ref{Hnumeq}) and (\ref{xnumeq}) are iterated until Eqs. (\ref{Hnumeq}) and (\ref{xnumeq}) converge. The corresponding resonator displacement and homodyne signals are plotted over the range of feedback phases in Fig. \ref{numerical scans}, for $\phi=\pi/2$ and  $\phi=0.4\pi$. We see that the results of the numerical model are similar to that of the analytical model, thus supporting its validity for our experiment.

\section{Experimental Results}
Having analysed the interference effects in the single-laser feedback cooling setup, we perform the experiment as described in the previous section, while fixing $\phi \simeq \pi/2$ using a servo controller and a piezo-mirror placed in the LO branch. The experimental setup is shown in Fig. (\ref{fig:setup}). Our experiments were performed with a silicon sliced photonic crystal nanobeam resonator, similar as has been presented in Refs. \onlinecite{leijssen_strong_2015,leijssen_nonlinear_2017,Shakespeare2021}, and a scanning electron microscope image of it is shown in the inset of Fig. \ref{fig:setup}. The main parameters at room temperature extracted in other experiments (data not shown) are mentioned in the caption. The sample is placed in a vacuum chamber which has a pressure of the order of $10^{-5}$ mbar. All experiments are performed at room temperature.

\begin{figure}
    \centering
    \includegraphics[width=0.5\textwidth]{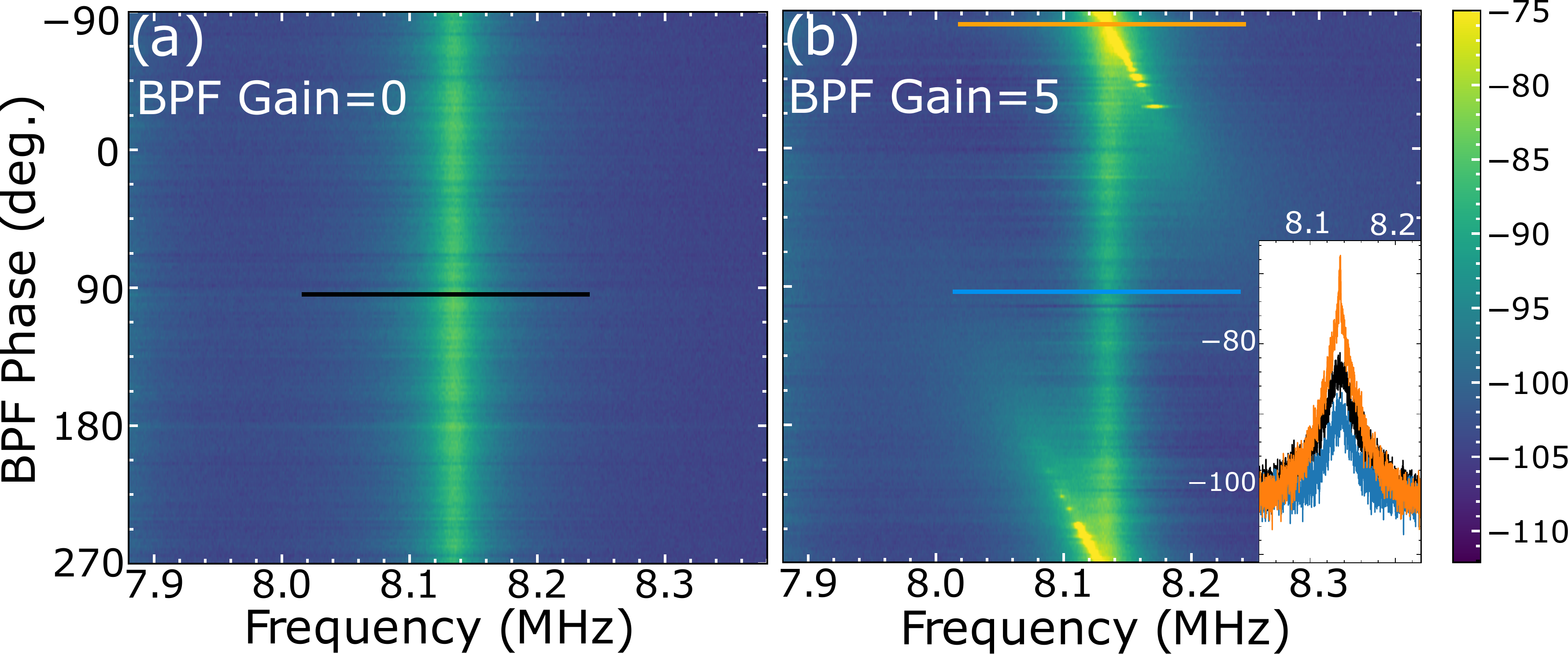}
    \caption{ Measured feedback phase dependence of homodyne signal spectrum for $\kappa=192$ GHz, $g_0=18$ MHz, $\Gamma=18$ kHz, $P_{in}=12$ \textmu W and applied feedback gains (a) BPF Gain$=0$ and (b) BPF Gain$=5$ for $\phi\simeq\pi/2$ using a BPF bandwidth of 77 kHz. The color bar (common to both plots - color-scale is over-saturated for (b)) represents the PSD of the measured homodyne signal in dBm. Inset: Line scans showing heating (orange) and cooling (blue) of resonator at different phases (marked in color plot) for BPF Gain = 5 compared to the spectrum at BPF Gain = 0 (black).}
    \label{expt scans}
\end{figure}

In Fig. \ref{expt scans}, we plot the experimental homodyne signal over the full range of feedback phases. We see that over the range of feedback gain shown here, the experimental results show similar regions of heating and cooling as seen in the numerical simulations and the analytical model for $\phi \simeq \pi/2$ (Figs. \ref{model scans} and \ref{numerical scans}), although there are some differences like a stronger signal around $\omega_M$ at all feedback phases. Given the similarities between these scans and the values of $\phi$ and BPF gain used in our experiment, we can infer that we are measuring in the parameter space ($\phi,\gamma_{fb}$) where the corresponding scans for the resonator displacement spectrum are expected to look similar to the homodyne signal scans (unlike in the case explored in Fig. \ref{stability}, for example).  When we apply a higher feedback gain, we see dominant features in the measured homodyne signal seemingly unrelated to the displacement of the resonator and possibly corresponding to the effects of the imprecision noise. This is plotted and explored in section C of SM.

Also, using Eq. (\ref{Hseq}), we can in principle determine the corresponding resonator displacement from the measured homodyne signal as the two are linearly related through the transduction parameter. In our case $\eta \approx 0.01$ and $\frac{\lvert a_{LO}\rvert}{\lvert a_{in}^0 \rvert} \approx 10$. From the control parameters of the piezo mirror (given in section D of SM), we estimate $\phi$ to be in the range of $0.49\pi-0.51 \pi$. Plugging these values into Eq. (\ref{transduction}), we see that, close to the optimal cooling point, the transduction parameter ratio $\beta/\beta_{\pi/2}$ is close to 1 (within 1 \%) for our experiment throughout the range of applied feedback gain (see section D of SM). Thus, we expect that the feedback gain dependence of the resonator displacement can be directly extracted by comparing the homodyne signal at various feedback gain values, allowing us to determine the extent of feedback cooling.

\begin{figure}
    \centering
    \includegraphics[width=0.5\textwidth]{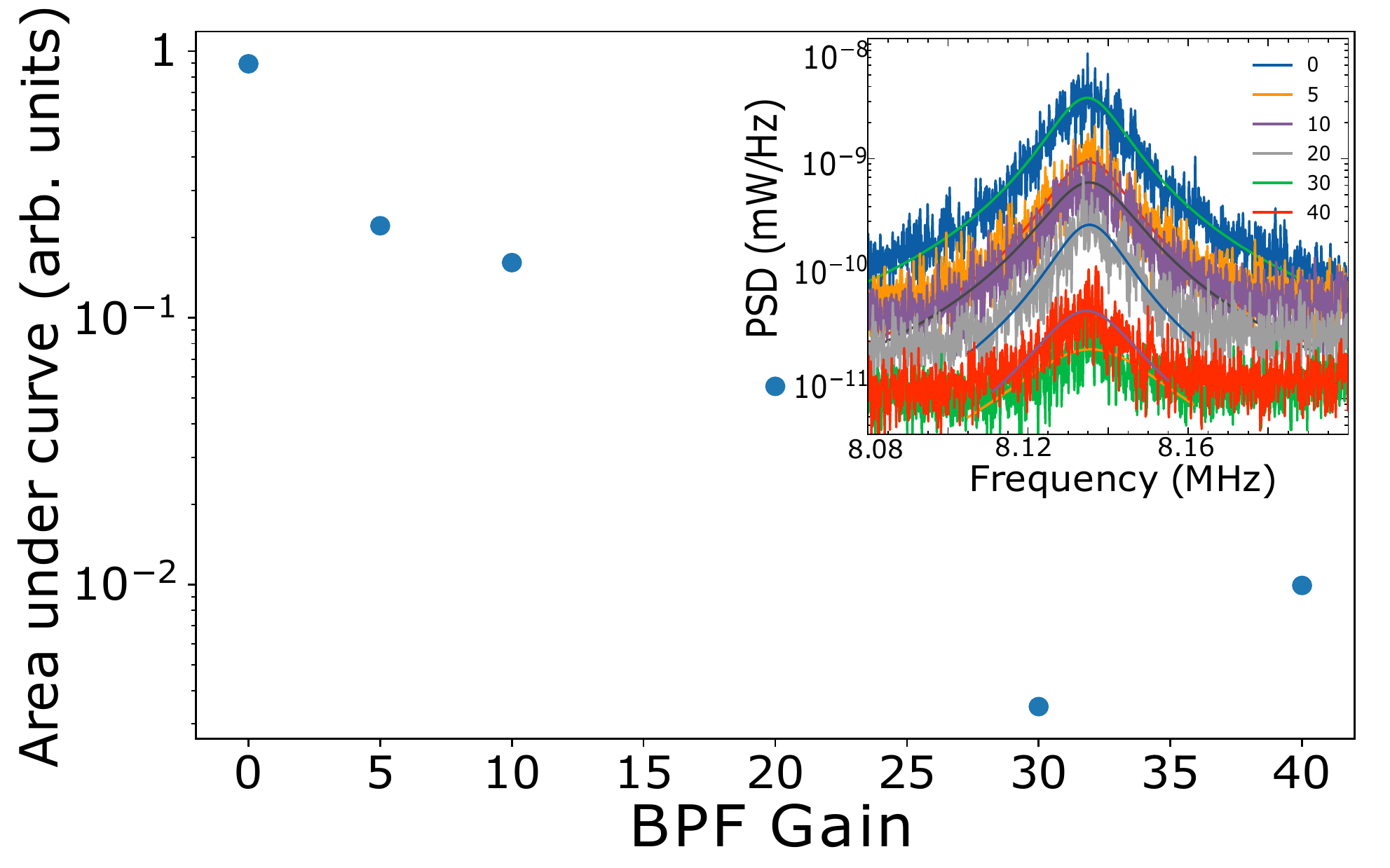}
    \caption{Area under measured spectrum of homodyne signal at the optimal cooling point (indicative of effective temperature of resonator) plotted against feedback gain. Corresponding spectra are shown in inset.}
    \label{cooling}
\end{figure}

In Fig. \ref{cooling}, we plot the spectra of the homodyne signal (and hence the scaled resonator displacement) at various gains at the optimal cooling point observed in the experiment. By comparing the area under these curves (which is proportional to the effective resonator temperature), we observe a cooling of more than two orders of magnitude before the effect of feedback gain on the noise floor presumably starts to heat up the resonator. 

When the homodyne angle deviates from $\pi/2$ by larger values than in our experiment however, the transduction parameter starts to depend more strongly on the applied feedback gain (Fig.~S4 in SM), and this has to be then accounted for while interpreting the homodyne signal, using the transduction parameter in the above model. This fact also indicates the importance of precise control of homodyne angle in such a single-laser feedback cooling experiment. 

Finally, while we have already demonstrated significant cooling of these silicon nanobeam resonators at room temperature with this technique, it is conceivable - from our analytical model - that higher cooling rates are possible to achieve when the homodyne angle is fixed further away from $\pi/2$, and with higher measurement efficiencies.

\section{Conclusions}

In summary, we have analytically, numerically and experimentally investigated a technique of single-laser feedback cooling as applied to silicon nanobeam resonators at room temperature. We find that, in this technique, it is important to consider the effect of modulations of laser beam used to control the resonator on the measurement itself. When the system is fixed at non-zero deviations from $\phi=\pi/2$, we find that this limits the feedback gain achievable in this technique beyond which the system becomes unstable. However, within the stable region, the homodyne signal can still be interpreted as a direct readout of the resonator displacement, but with a feedback gain and phase dependant transduction parameter. Further, we find that this interference effect can be used to enhance the cooling of the resonator at feedback gain dependant homodyne angles away from the conventional $\pi/2$ value. We further identify regions of heating and cooling of the resonator in the homodyne signal spectrum as a function of the feedback phase, when the homodyne angle is fixed close enough to $\pi/2$ that these features seen in the homodyne signal can be directly interpreted as resulting from the resonator displacement. This is further verified through numerical simulation and experimental demonstration. 

Our study thus introduces important interference effects to be considered when the same laser beam is used for measurement and control in a feedback cooling setup. It also proposes a new optimal homodyne angle that considers these interference effects where the cooling effects can be stronger than in the conventional auxiliary laser case.

\begin{acknowledgments}
We acknowledge Amy Navarathna and Ewold Verhagen for providing us the measured sample. This project has received funding from the European Research Council (ERC) under the European Union’s Horizon 2020 research and innovation programme (grant agreement No 852428) and from Academy of Finland Grant No 321416.
\end{acknowledgments}

\bibliography{single_refs}

\begin{thebibliography}{30}
\bibitem [{not()}]{note1}%
  \BibitemOpen
  \href@noop {} {}\bibinfo {note} {Note that in our cross-polarized setup
  $a_{in}$ and $a_s$ have opposite polarizations and hence there is no
  interference between them. But even if there was, using the conventional
  input-output form $a_s=a_{in}-\sqrt{\kappa}a$ would not change the results we
  present below as long as care is taken in defining the homodyne angle
  $\phi$.}\BibitemShut {Stop}%
\end{thebibliography}

\clearpage
\onecolumngrid

\renewcommand{\thefigure}{S\arabic{figure}}
\renewcommand{\thetable}{S\arabic{table}}
\renewcommand{\theequation}{S\arabic{equation}}

\setcounter{figure}{0}
\setcounter{equation}{0}

\section*{Supplementary Information}

\subsection{Analytical Model Details}
In the non-resolved sideband regime, where the optical field mode $a$ reaches a steady state much faster than the mechanical mode, the output signal of the homodyne interferometer (in the frame rotating at the laser frequency), within the setup described above, is given by Eq. (\ref{model_SM})
\begin{equation}
    H_0=\lvert a_{LO}\rvert \lvert a_{in}^0\rvert 4 \eta\frac{\cos{\phi}+\frac{2\Delta}{\kappa}\sin{\phi}}{1+(\frac{2\Delta}{\kappa})^2}
\end{equation}
where $\eta=\kappa_{ex}/\kappa$, and the effective detuning $\Delta=\Delta_0 + G x(t)$, where $G$ is the single-photon optomechanical coupling rate, $\Delta_0$ is the cavity detuning and $x(t)$  is the displacement of the resonator.

For small displacements of the resonator where $\frac{2\Delta}{\kappa} \ll 1$, Eq. (\ref{model_SM}) reduces to 
\begin{equation}
    H_0=\lvert a_{LO}\rvert \lvert a_{in}^0\rvert 4 \eta(\cos{\phi}+\frac{2\Delta}{\kappa}\sin{\phi})
\end{equation}
This signal is then passed into the bandpass filter, where only frequencies around $\omega_M$ are allowed to pass through with a gain of $\gamma_{fb}$ and an additional phase factor of $\theta_{fb}$, which therefore selects terms that are linear in $x(t)$ (since $\Delta_0$ is a constant and the higher power terms have frequencies corresponding to $2\Delta$,$3\Delta$ etc ) gives us an EOM input signal given by 
\begin{equation}
 H_{EOM}^0  = \lvert a_{LO}\rvert \lvert a_{in}^0 \rvert 4 \eta(\frac{2Gx_0(t) e^{i\theta_{fb}}}{\kappa}\sin{\phi})\gamma_{fb} 
\end{equation}

The EOM modulates the input signal intensity around the average intensity $\lvert a_{in}^0 \rvert^2$.  The new input signal that then strikes the sample is given by

\begin{equation}
\lvert a_{in}^1 \rvert^2 =\lvert a_{in}^0 \rvert^2(1 + \gamma_c H_{EOM}^{0'}) =\vert a_{in}^0 \rvert^2(1+ \gamma_c \frac{2Gx_0(t)}{\kappa}e^{i\theta_{fb}}\sin{\phi}\gamma_{fb} )
\end{equation}
where $\gamma_c=\frac{V_{detector}}{V_{EOM}}(\frac{\lvert a_{LO}\rvert}{\lvert a_{in}^0 \rvert}4 \eta)$ where  $V_{detector}$ and ${V_{EOM}}$ are the detector and EOM coversion parameters respectively that correspond to converting the optical power to voltage and vice versa, and $H_{EOM}^{0'}=H_{EOM}^{0}/4\eta\lvert a_{LO}\rvert \lvert a_{in}^0\rvert$.

The homodyne signal that is a result of this input signal is therefore given by (from Equation S2) 
\begin{equation}
    H_1=\lvert a_{LO}\rvert \lvert a_{in}^0\rvert \sqrt{1+ \gamma_c \frac{2Gx_0(t)}{\kappa}e^{i\theta_{fb}}\sin{\phi}\gamma_{fb}  } 4 \eta(\cos{\phi}+\frac{2\Delta_{x_1(t)}}{\kappa}\sin{\phi})
\end{equation}
Where $x_1(t)$ is the the resonator displacement due to the modulated input $a_{in}^1$. Taylor expanding gives
\begin{equation}
    H_1=\lvert a_{LO}\rvert \lvert a_{in}^0\rvert ( 1+\frac{1}{2}\gamma_c \frac{2Gx_0(t)}{\kappa}e^{i\theta_{fb}}\sin{\phi}\gamma_{fb}  -..)  4 \eta(\cos{\phi}+\frac{2\Delta_{x_1(t)} }{\kappa}\sin{\phi})
\end{equation}
When this is fed back into the EOM after the band pass filter, it gives us 
\begin{equation}
\begin{split}
 H_{EOM}^1  = \lvert a_{LO}\rvert \lvert a_{in}^0 \rvert [4 \eta(\frac{2Gx_1(t) e^{i\theta_{fb}}}{\kappa}\sin{\phi})\gamma_{fb}  \\ + \frac{1}{2}   (4 \eta) \gamma_c \frac{2Gx_0(t)}{\kappa} e^{2i\theta_{fb}}\sin{\phi}\gamma_{fb}  \cos{\phi}\gamma_{fb} ] \\
= \lvert a_{LO}\rvert \lvert a_{in}^0 \rvert 4 \eta( e^{i\theta_{fb}}\sin{\phi})\gamma_{fb} [\frac{2Gx_1(t)}{\kappa}+\frac{2Gx_0(t)}{\kappa}\frac{1}{2}\gamma_c e^{i\theta_{fb}}\cos{\phi}\gamma_{fb}]
\end{split}
\end{equation}
Therefore, the input signal for the second time step is given by
\begin{equation}
\begin{split}
    \lvert a_{in}^2 \rvert^2=\lvert a_{in}^0 \rvert^2(1 + \gamma_c H_{EOM}^{1'}) =\vert a_{in}^0 \rvert^2(1+ \gamma_c e^{i\theta_{fb}}\sin{\phi}\gamma_{fb} \\ \times[\frac{2Gx_1(t)}{\kappa}+\frac{2Gx_0(t)}{\kappa}\frac{1}{2}\gamma_c e^{i\theta_{fb}}\cos{\phi}\gamma_{fb}])
\end{split}
\end{equation}
Repeating the process to derive the homodyne signal and the input signal to the EOM for the second time step gives us
\begin{equation}
\begin{split}
    H_2=\lvert a_{LO}\rvert \lvert a_{in}^0\rvert \sqrt{1+ \gamma_c e^{i\theta_{fb}}\sin{\phi}\gamma_{fb}[\frac{2Gx_1(t)}{\kappa}+\frac{2Gx_0(t)}{\kappa}\frac{1}{2}\gamma_c e^{i\theta_{fb}}\cos{\phi}\gamma_{fb}]} \\\times 4 \eta(\cos{\phi}+\frac{2\Delta_{x_2(t)}}{\kappa}\sin{\phi})\\
    =\lvert a_{LO}\rvert \lvert a_{in}^0\rvert (1+\frac{1}{2}\gamma_c e^{i\theta_{fb}}\sin{\phi}\gamma_{fb}[\frac{2Gx_1(t)}{\kappa}+\frac{2Gx_0(t)}{\kappa}\frac{1}{2}\gamma_c e^{i\theta_{fb}}\cos{\phi}\gamma_{fb}]+...)\\\times 4 \eta(\cos{\phi}+\frac{2\Delta_{x_2(t)}}{\kappa}\sin{\phi})
\end{split}
\end{equation}

\begin{equation}
\begin{split}
 H_{EOM}^2  =  \lvert a_{LO}\rvert \lvert a_{in}^0 \rvert [4 \eta(\frac{2(Gx_2(t)) e^{i\theta_{fb}}}{\kappa}\sin{\phi})\gamma_{fb}\\ + \frac{1}{2}\gamma_c e^{2i\theta_{fb}}\sin{\phi}\gamma_{fb}^2[\frac{2Gx_1(t)}{\kappa}+\frac{2Gx_0(t)}{\kappa}\frac{1}{2}\gamma_c e^{i\theta_{fb}}\cos{\phi}\gamma_{fb}]4 \eta\cos{\phi}]\\
=\lvert a_{LO}\rvert \lvert a_{in}^0 \rvert 4 \eta( e^{i\theta_{fb}}\sin{\phi})\gamma_{fb}(\frac{2Gx_2(t)}{\kappa}+\frac{1}{2}\gamma_c e^{i\theta_{fb}}\gamma_{fb}\cos{\phi}[\frac{2Gx_1(t)}{\kappa}+\frac{2Gx_0(t)}{\kappa}\frac{1}{2}\gamma_c e^{i\theta_{fb}}\cos{\phi}\gamma_{fb}])\\
= \lvert a_{LO}\rvert \lvert a_{in}^0 \rvert 4 \eta( e^{i\theta_{fb}}\sin{\phi})\gamma_{fb}(\frac{2Gx_2(t)}{\kappa}+\frac{2Gx_1(t)}{\kappa}\frac{1}{2}\gamma_c e^{i\theta_{fb}}\gamma_{fb}\cos{\phi}+\frac{2Gx_0(t)}{\kappa}(\frac{1}{2}\gamma_c e^{i\theta_{fb}}\gamma_{fb}\cos{\phi})^2)
\end{split}
\end{equation}

Therefore, after $N$ time steps

\begin{equation}
\begin{split}
 H_{EOM}^{N}  =  
 \lvert a_{LO}\rvert \lvert a_{in}^0 \rvert 4 \eta( e^{i\theta_{fb}}\sin{\phi})\gamma_{fb}(\frac{2Gx_N(t)}{\kappa}+\frac{2Gx_{N-1}(t)}{\kappa}\frac{1}{2}\gamma_c e^{i\theta_{fb}}\gamma_{fb}\cos{\phi}\\+\frac{2Gx_{N-2}(t)}{\kappa}(\frac{1}{2}\gamma_c e^{i\theta_{fb}}\gamma_{fb}\cos{\phi})^2+...+\frac{2Gx_1(t)}{\kappa}(\frac{1}{2}\gamma_c e^{i\theta_{fb}}\gamma_{fb}\cos{\phi})^{N-1}+\frac{2Gx_0(t)}{\kappa}(\frac{1}{2}\gamma_c e^{i\theta_{fb}}\gamma_{fb}\cos{\phi})^N)
 \end{split}   
\end{equation}

Around the frequency of interest $\omega_M$, the homodyne signal after N steps (before adding gain and phase factor for the previous equation) is
\begin{equation}
\begin{split}
    H_N^{\omega\sim \omega_M}=
    \lvert a_{LO}\rvert \lvert a_{in}^0 \rvert 4 \eta( \sin{\phi})(\frac{2Gx_N(t)}{\kappa}+\frac{2Gx_{N-1}(t)}{\kappa}\frac{1}{2}\gamma_c e^{i\theta_{fb}}\gamma_{fb}\cos{\phi}\\+\frac{2Gx_{N-2}(t)}{\kappa}(\frac{1}{2}\gamma_c e^{i\theta_{fb}}\gamma_{fb}\cos{\phi})^2+...+\frac{2Gx_1(t)}{\kappa}(\frac{1}{2}\gamma_c e^{i\theta_{fb}}\gamma_{fb}\cos{\phi})^{N-1}+\frac{2Gx_0(t)}{\kappa}(\frac{1}{2}\gamma_c e^{i\theta_{fb}}\gamma_{fb}\cos{\phi})^N)
\end{split}
\end{equation}

To summarize, the final feedback signal that goes to the EOM is given by 
\begin{equation}
 H_{EOM}  =  \lvert a_{LO}\rvert \lvert a_{in}^0 \rvert 4 \eta( e^{i\theta_{fb}}\sin{\phi})\gamma_{fb}\frac{2G}{\kappa}\lim_{N\to\infty} [\sum_{n=0}^{N}x_{N-n}(t)(\frac{1}{2}\gamma_c e^{i\theta_{fb}}\gamma_{fb}\cos{\phi})^n]
 \label{EOMeqSM}
\end{equation}

and the resultant Homodyne signal around $\omega\simeq \omega_M$ is 
\begin{equation}
    H^{\omega\sim \omega_M}=\lvert a_{LO}\rvert \lvert a_{in}^0 \rvert 4 \eta( \sin{\phi})\frac{2G}{\kappa}\lim_{N\to\infty} [\sum_{n=0}^{N}x_{N-n}(t)(\frac{1}{2}\gamma_c e^{i\theta_{fb}}\gamma_{fb}\cos{\phi})^n]
    \label{HeqSM}
\end{equation}

For small enough gains, where the series in Eq. (\ref{EOMeqSM}) is convergent, we expect that the resonator displacement eventually converges to $x_s(t)$. Therefore, Eq. (\ref{EOMeqSM}) gives us an input signal
\begin{equation}
    \lvert a_{in}^s \rvert \simeq \lvert a_{in}^0 \rvert(1+\frac{1}{2}\gamma_c e^{i\theta_{fb}}\sin{\phi}\gamma_{fb} \frac{2G }{\kappa} x_{s}(t)[\sum_{n=0}^{\infty}(\frac{1}{2}\gamma_c e^{i\theta_{fb}}\gamma_{fb}\cos{\phi})^n])   
\label{anseries}
\end{equation}
 since removing a finite number of terms from the end of a convergent infinite series does not change its value.
 
We see that the input term is directly proportional to $x_s$, the displacement of the oscillator. This corresponds to feedback control, which can be used to heat/cool the resonator depending on the applied feedback phase.

The corresponding Homodyne signal around $\omega_M$,
\begin{equation}
    H_s(\omega)^{\omega\sim\omega_M} = \lvert a_{in}^0 \rvert\lvert a_{LO}\rvert 4 \eta\sin{\phi}y(\omega)[\sum_{n=0}^{\infty}(\frac{1}{2}\gamma_c e^{i\theta_{fb}}\gamma_{fb}\cos{\phi})^n]
    \label{Hseqn}
\end{equation}
where $y(\omega)=\frac{2G }{\kappa}x_s(\omega)$ is the rescaled resonator displacement.

The corresponding Power Spectral Density is

\begin{equation}
    S_{HH}(\omega)^{\omega\sim\omega_M} =\left\vert \lvert a_{in}^0 \rvert\lvert a_{LO}\rvert 4 \eta\sin{\phi}[\sum_{n=0}^{\infty}(\frac{1}{2}\gamma_c e^{i\theta_{fb}}\gamma_{fb}\cos{\phi})^n] \right\vert^2 S_{yy}(\omega) 
    \label{SHHseqn}
\end{equation}

\subsection{Resonator displacement from analytical model}
\begin{figure*}[h]
    \centering
    \renewcommand{\thefigure}{S1}
    \includegraphics[width=\textwidth]{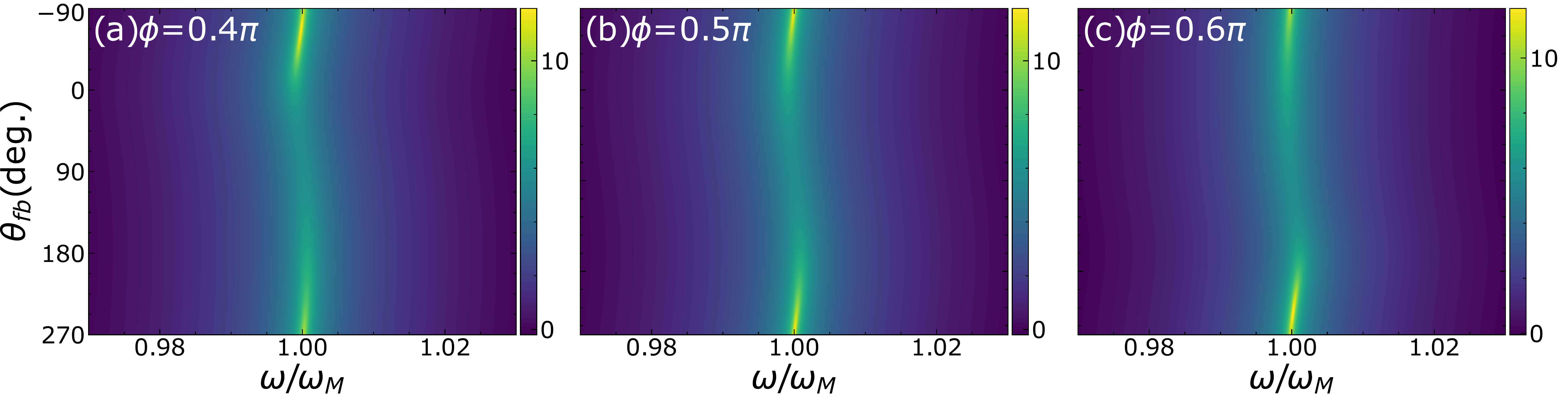}
    \caption{Resonator displacement spectra calculated from analytical model plotted over the range of feedback phases for the system parameters given in the caption of Fig. 2 in main text and $\gamma_{fb}=7$, for $\phi=0.4\pi, \pi/2, 0.6 \pi$. The color bar represents the PSD of the calculated homodyne signal in an arbitrary logarithmic scale. }
    \label{model_SM}
    
\end{figure*}
From Fig. \ref{model_SM}, we see that the feedback phase dependence scans for the resonator displacement calculated from the analytical model are similar to the homodyne signal scans shown in Fig. 2 of the main text for ($\phi$,$\gamma_{fb}$) in the given range.

\subsection{Experimental feedback phase scans at higher feedback gain}
\begin{figure*}[h]
    \centering
    \renewcommand{\thefigure}{S2}
    \includegraphics[width=\textwidth]{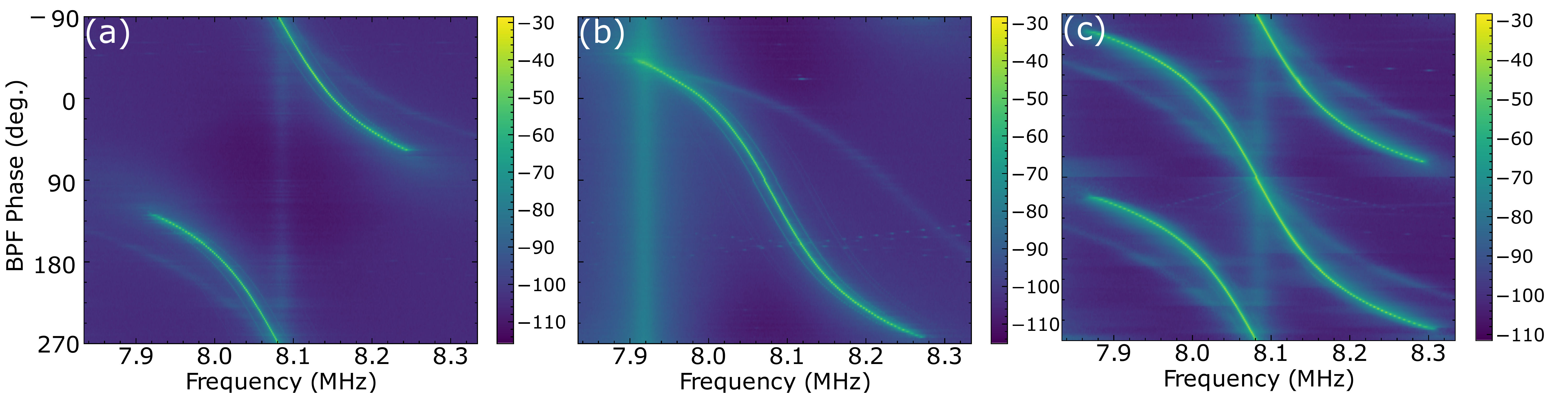}
    \caption{Measured homodyne signal plotted over the range of feedback phases for a feedback gain of 30 for different conditions:  (a) Piezo mirror fixed at $\phi\simeq\pi/2$ and BPF center frequency tuned to mechanical resonance frequency of the resonator (b) Piezo mirror fixed at $\phi\simeq\pi/2$ and BPF center frequency offset 70 kHz higher than mechanical frequency of the resonator (BPF bandwidth= 77 kHz)  (c) Piezo mirror sweeping $\phi$ from 0 to 2$\pi$ at 100 Hz and BPF center frequency tuned to mechanical resonance frequency of the resonator.}
    \label{expt_SM}
\end{figure*}

At a BPF gain of 30, from Fig. \ref{expt_SM}(a), we observe bright lines originating around BPF phase= 90 degrees, which extend outwards and curve away from the BPF center frequency. In Fig. \ref{expt_SM}(b), we offset the BPF center frequency away from the mechanical resonance frequency. In this case, we still observe that the lines originate from the BPF center frequency and the spectra around the mechanical resonance frequency remains largely unaffected. We can therefore conclude that these lines are not directly related to the resonator dynamics. In Fig. \ref{expt_SM}(c), we sweep $\phi$ from 0 to $2\pi$ at 100 Hz instead of fixing it at $\pi/2$, and we observe an additional bright line that originates around BPF phase = -90 degrees.

\subsection{Piezo control and transduction parameter}
\begin{figure*}[h]
    \centering
    \renewcommand{\thefigure}{S3}
    \includegraphics[width=0.6\textwidth]{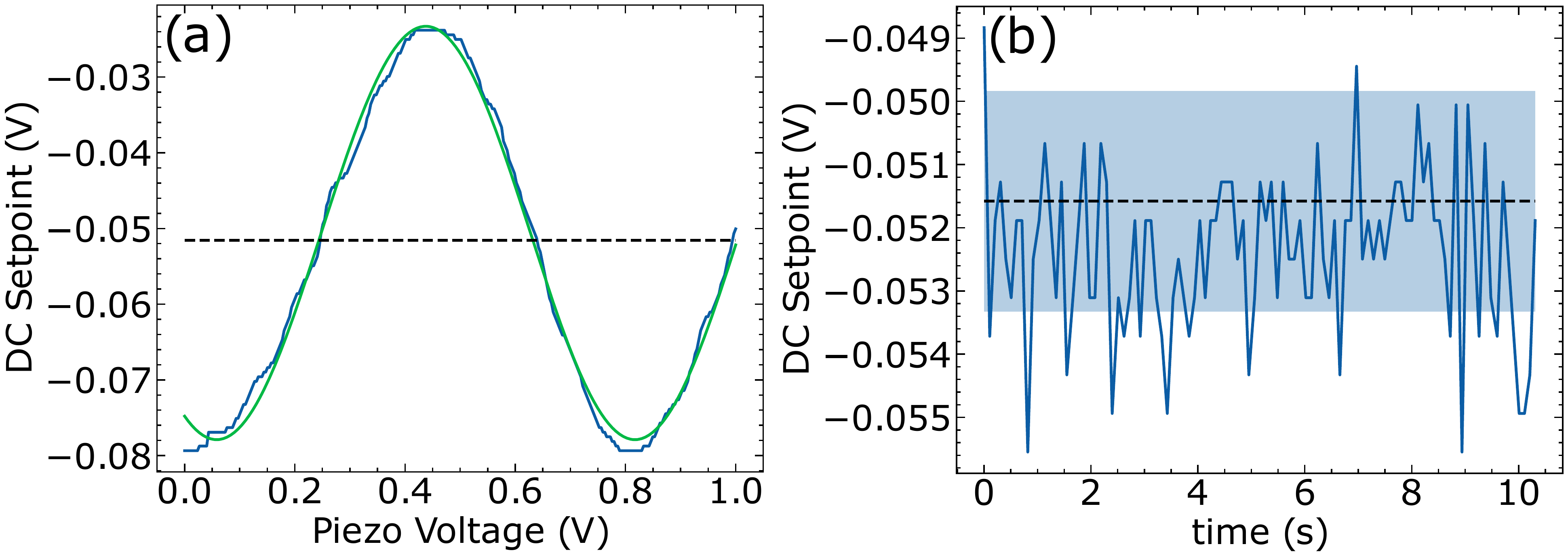}
    \caption{(a)Piezo voltage sweep curve. Blue line shows the DC component of the output signal from the homodyne detector (DC setpoint) versus the applied voltage to the piezo mirror. Green line shows fit to cosine dependence. Dashed line marks the DC setpoint corresponding to $\phi=\pi/2$. (b) DC setpoint fixed at $\phi\simeq \pi/2$ using servo controller, plotted versus time. Dashed line marks the DC setpoint corresponding to $\phi=\pi/2$ and borders of blue shaded region mark DC setpoints corresponding to $\phi=0.49 \pi$ and $\phi=0.51 \pi$}
    \label{piezo_SM}
    
\end{figure*}

\begin{figure*}
    \centering
    \includegraphics[width=0.6\textwidth]{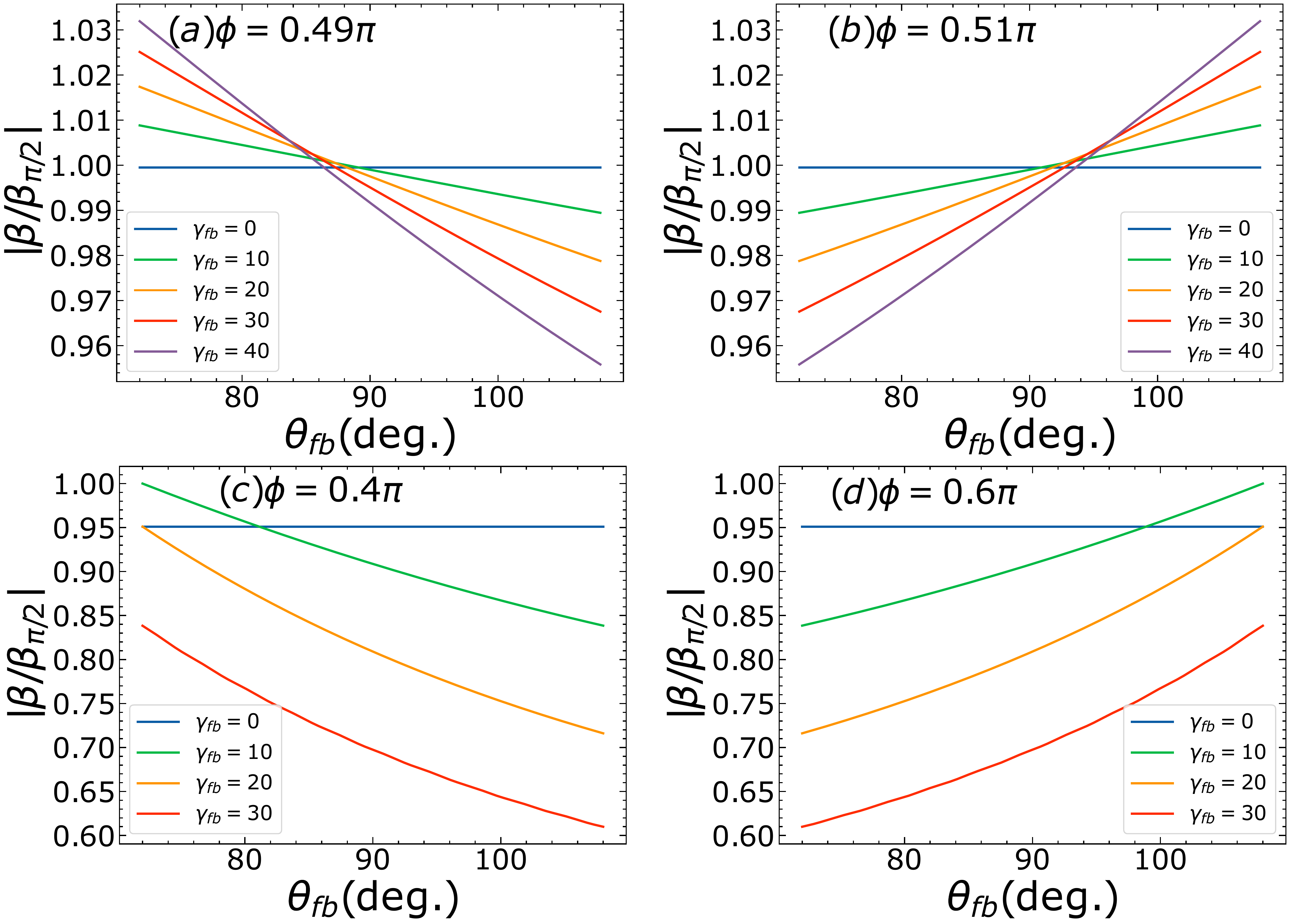}
    \caption{Transduction parameter ratio calculated from analytical model plotted versus feedback phase, close to the optimal cooling point  }
    \label{trans_SM}
\end{figure*}

In Fig. \ref{piezo_SM}, we show the details of locking the homodyne angle at $\phi\simeq\pi/2$ for our experiment, and estimate the corresponding fluctuation in $\phi$. In Fig. \ref{trans_SM}, we show the transduction parameter between homodyne signal and resonator displacement (Eq.(10) in main text) as a function of feedback phase for various feedback gain, both at the homodyne angles (close to $\pi/2$) used in the experiment and at larger deviations from $\phi=\pi/2$.  

\end{document}